\documentclass[runningheads]{llncs}

\usepackage[utf8]{inputenc}
\usepackage[T1]{fontenc}
\usepackage{graphicx}
\usepackage{booktabs}
\usepackage{array}
\usepackage{multirow}
\usepackage{amsmath}
\usepackage{float}
\usepackage{placeins}
\usepackage{url}
\usepackage{color}
\usepackage{hyperref}
\usepackage{cite}

\newcommand{\includegraphicsifexists}[2][]{%
  \IfFileExists{#2}{\includegraphics[#1]{#2}}{%
    \fbox{\begin{minipage}[c][0.22\textheight][c]{0.92\linewidth}
    \centering Missing figure file: \texttt{\detokenize{#2}}
    \end{minipage}}%
  }%
}

\title{Benchmarking Literature Retrieval for a Model Organism: A \emph{Dictyostelium} Case Study}
\titlerunning{Model-Organism Literature Retrieval: \emph{Dictyostelium}}

\author{Yun Wang\inst{1}\orcidID{0000-0002-2441-4182} \and
Gad Shaulsky\inst{2}\orcidID{0000-0002-0532-0551} \and
Toma\v{z} Curk\inst{1}\orcidID{0000-0003-4888-7256} \and
Bla\v{z} Zupan\inst{1,2}\orcidID{0000-0002-5864-7056}}

\authorrunning{Y. Wang et al.}

\institute{
University of Ljubljana, Faculty of Computer and Information Science,
Ve\v{c}na pot 113, 1000 Ljubljana, Slovenia\\
\email{\{yun.wang,tomaz.curk,blaz.zupan\}@fri.uni-lj.si}
\and
Department of Molecular and Human Genetics, Baylor College of Medicine,
Baylor Plaza, Houston, TX 77030, USA\\
\email{gadi@bmc.edu}
}
\begin{document}

\maketitle
\emergencystretch=3em

\begin{abstract}
Biological literature retrieval systems are often developed and evaluated using broad biomedical corpora and general-purpose search tasks. However, many curated knowledge bases operate in narrower model-organism domains, where the literature is sparse and terminology is organism-specific. We introduce a retrieval benchmark from dictyBase for \emph{Dictyostelium}, a model organism in cell and developmental biology. The benchmark consists of curator-generated biological queries linked to PubMed-indexed articles, together with structured gene annotations. Using this benchmark, we study three factors in niche biological retrieval: cross-encoder reranking, gene-aware query expansion, and abstract-only versus full-text retrieval. We report that reranking and gene-aware query expansion improve retrieval selectively: reranking is most useful when the model is well suited to biological evidence matching, whereas curated annotations help clarify compact biological queries by reducing vocabulary mismatch. Full-text chunks substantially improve retrieval when abstracts omit supporting evidence, increasing both candidate recall and top-rank performance, although these cases are harder than queries supported by abstracts. Data and code are publicly available at \url{https://github.com/fulaibaowang/dictycite}, and the benchmark dataset is additionally archived on Zenodo.

\end{abstract}

\section{Introduction}

Biological literature retrieval systems are usually developed and benchmarked
on broad biomedical corpora and general search tasks. However, a large amount of biological
knowledge is maintained in narrower model-organism communities, such as
\emph{D. discoideum}, \emph{D. melanogaster}, \emph{C. elegans}. In these settings, the literature is
smaller, terminology is more organism-specific, and relevant evidence may depend
on gene names, aliases, product annotations, and phenotype terms rather than on
broadly shared biomedical vocabulary. Retrieval behavior in such domains may
therefore differ from what is observed in large-scale biomedical question answering
or general literature search.

Such differences raise a practical question for scientific retrieval systems: how well do standard retrieval pipelines work when given a compact biological statement and asked to retrieve a proper supporting reference from a curated knowledge base? Systems that perform well on broad literature-search
benchmarks may behave differently in narrower biological domains, where exact
gene names, aliases, and phenotype terms can carry substantial signal. In this
setting, we evaluate two common strategies: semantic reranking, which reorders the retrieved articles using a selective model, and query expansion, which adds related terms to the search query.

Many model-organism databases maintain gene-centered pages that combine
structured annotations, textual summaries, and links to supporting literature~\cite{ozturk-colak_flybase_2024,harris_wormbase_2020}. Among these resources, dictyBase covers \emph{Dictyostelium}, a social amoeba
widely used in biology, including studies of chemotaxis,
cell motility, phagocytosis, and multicellular development.
dictyBase is especially useful for studying retrieval in this setting because its
gene summary pages contain curator-generated notes with inline citations. This
structure allows the curation record to be converted into a controlled evidence
retrieval task: given a compact biological statement, retrieve a proper citation to an
article from a domain-specific literature corpus.

In this paper, we construct a citation-derived retrieval benchmark from
dictyBase~\cite{fey_one_2013} and use it to study literature retrieval for
\emph{Dictyostelium}. We use this setting to examine three questions: whether reranking adds value beyond strong keyword-based retrieval, whether curated gene annotations
help compact biological queries, and how full-text evidence changes retrieval
when abstracts are insufficient.

Our principal contribution is not a new retrieval architecture, but a benchmark
and analysis of retrieval behavior in a sparse model-organism domain. The
results show that reranking gains depend strongly on model fit, that trusted
gene annotations can reduce vocabulary mismatch, and that full-text chunks improve
recall and final ranking when abstracts omit detailed evidence, while not fully
closing the gap to claims whose evidence is already present in abstracts.

\section{Related Work}
Biomedical literature retrieval has been studied through broad benchmarks and shared tasks such as BioASQ~\cite{tsatsaronis_overview_2015} and the TREC BioGen track~\cite{coordinators-trec2025-papers-proc-4}, as well as through curation-oriented search and annotation systems such as Textpresso~\cite{muller_textpresso_2004} and PubTator~\cite{wei_pubtator_2013}. These efforts cover biomedical question-answering, semantic indexing, text mining, and literature search support. Our work is related in motivation, but focuses on a different evaluation setting: retrieving supporting citations for compact, detailed, domain-specific biological queries, where terminology, synonyms, and curated annotations may affect retrieval differently from broad biomedical QA settings.

Modern retrieval systems commonly use two-stage pipelines, combining efficient first-stage retrieval with more expensive reranking. Lexical retrieval~\cite{robertson_probabilistic_2009}, dense retrieval~\cite{li_making_2023}, reciprocal rank fusion~\cite{cormack_reciprocal_2009}, cross-encoder reranking~\cite{bajaj_ms_2018}, and LLM-based rerankers are widely used components.
Biomedical rerankers such as MedCPT further show the value of domain adaptation for PubMed-scale retrieval~\cite{jin_medcpt_2023}. We use these standard components to examine how modern retrieval pipelines behave in a niche scientific retrieval setting.

Query expansion and full-text retrieval are long-standing topics in biomedical search. Prior work has used controlled vocabularies, MeSH terms, gene synonyms, and concept-based expansion to reduce vocabulary mismatch, while full-text retrieval can recover evidence absent from abstracts~\cite{lu_evaluation_2009,lu_empirical_2009,kim_towards_2022}. We study both factors in a controlled setting by using curated annotations for query expansion and by comparing abstract-only retrieval with chunked full-text retrieval.

Taken together, prior work provides strong retrieval components and broad
biomedical retrieval benchmarks, but less is known about their behavior in model-organism knowledge bases. Our work uses dictyBase
claim--citation pairs as a controlled test case for this setting, with emphasis on
reranker domain fit, curated gene annotations, and full-text evidence.

\section{Dataset}
\label{sec:dataset}

We constructed a \emph{Dictyostelium} literature retrieval dataset from dictyBase gene summary pages.
Each retrieval instance is derived from a curator-generated biological claim and its inline literature citation: the claim becomes the query, and the cited PubMed-indexed articles are used as the relevance judgment. 
The dataset contains 1,656 curator-claim queries from 862 unique \emph{Dictyostelium} genes, linked to 1,289 unique PubMed articles and evaluated against a 20,447-article Europe PMC abstract corpus. 
In addition to the query--article relevance pairs, the dataset includes structured gene annotations for gene-aware query expansion and LLM-assisted evidence labels for stratified analysis of whether the cited abstract contains sufficient evidence for the curator claim. 
The dataset is available on Zenodo under DOI \href{https://doi.org/10.5281/zenodo.20308282}{10.5281/zenodo.20308282}.

\begin{figure}[!htbp]
\centering
\includegraphicsifexists[width=0.98\linewidth,height=0.48\textheight,keepaspectratio]{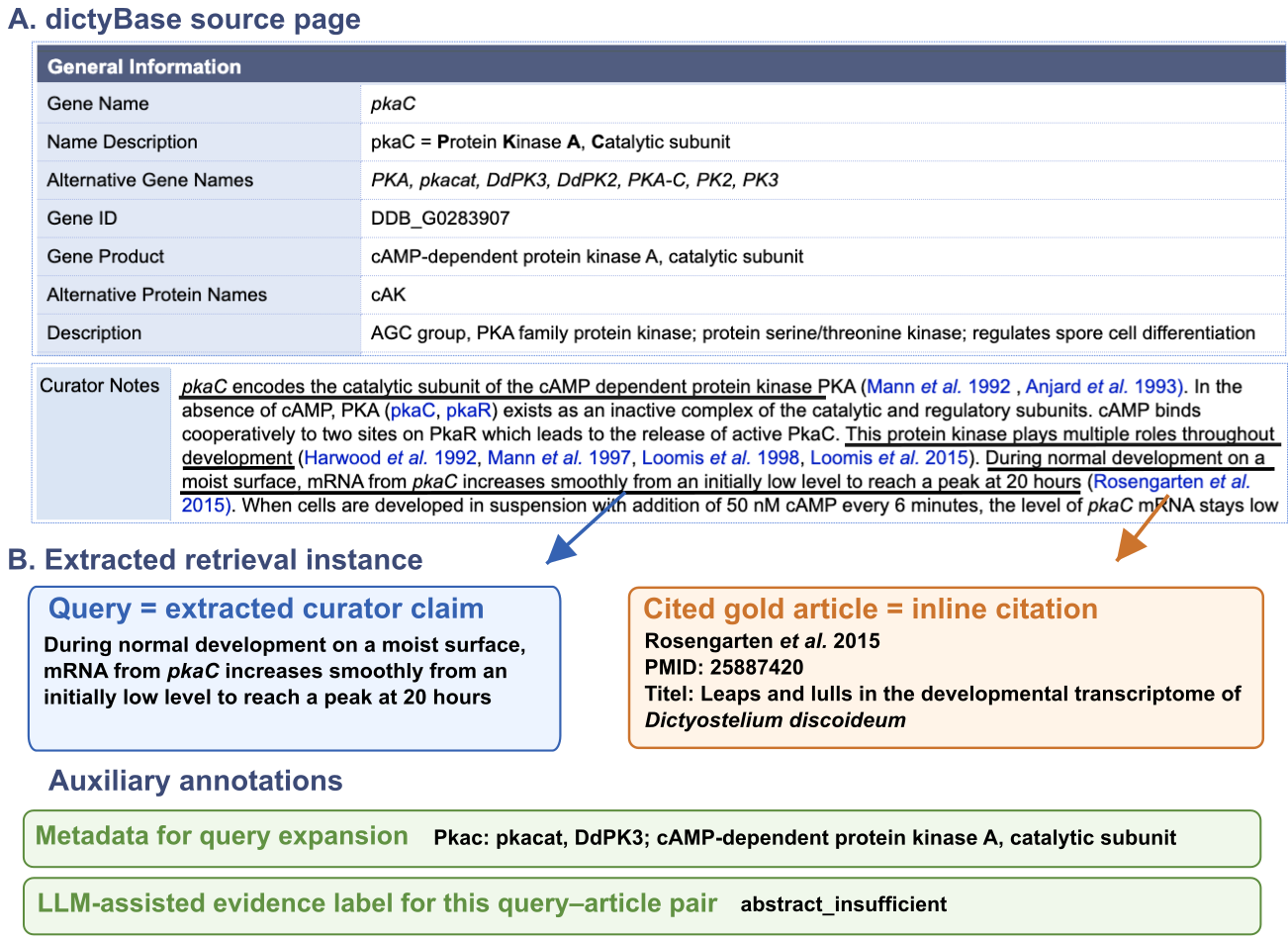}
\caption{Construction of a retrieval instance from a dictyBase gene page. 
(A) The source page for gene \emph{pkaC} 
(\url{http://dictybase.org/gene/DDB_G0283907}) contains structured gene
annotations and curator-generated notes with inline citations. 
(B) A curator-generated claim is extracted as the query, and its inline citation 
is mapped to a PubMed-indexed article used as the cited relevance judgment. 
Auxiliary fields are used in later experiments: gene annotations support 
gene-aware query expansion, and the LLM-assisted evidence label supports the 
full-text analysis by marking whether the cited abstract is sufficient for the 
claim or whether full-text evidence may be needed.}
\label{fig:dataset}
\end{figure}

\paragraph{Retrieval corpus.}
We assembled the retrieval corpus by collecting PubMed-indexed articles related to \emph{Dictyostelium} from Europe PMC. 
For the abstract-based experiments, we used the 20,447 records with available abstracts, which were cleaned by removing formatting tags and normalizing whitespace, resulting in a collection of 20,447 articles. 
All abstract-only retrieval and reranking experiments use this same corpus.

\paragraph{Curator-derived queries and relevance judgments.}
Queries were derived from curator notes on dictyBase gene summary pages. 
These notes consist of compact biological statements supported by inline citations, as illustrated in Fig.~\ref{fig:dataset}. 
For example, the statement that during normal development on a moist surface, \emph{pkaC} mRNA increases from an initially low level to a peak at 20 hours is extracted as a retrieval query, and the cited Rosengarten \emph{et al.} article is used as the reference article for evaluation.
Because dictyBase uses internal publication identifiers, we first mapped cited publications to PubMed IDs. 
Claims were excluded when their citations could not be mapped to PubMed or when the cited articles were absent from the Europe PMC abstract corpus, removing about 2\% of the claims in total.
The remaining claims were cleaned and deduplicated. 
Near-duplicate or semantically similar claims were grouped within each gene, and one representative claim was retained as the final query. 

\paragraph{Gene annotations for query expansion.}
For each gene page, dictyBase provides structured annotations that connect a canonical gene name to alternative gene names, protein names, and product descriptions. These annotations are useful because relevant papers do not always use the canonical gene name. For example, \emph{pkaC} (Fig.~\ref{fig:dataset}A) encodes PKA-C, the catalytic subunit of cAMP-dependent protein kinase A, where cAMP stands for cyclic adenosine monophosphate. A paper may therefore refer to the gene, the protein, the kinase activity, or cAMP without using \emph{pkaC} itself.
We retain these fields for query expansion only, not as part of the original query.
They are used only in the gene-aware query-expansion experiments described in Section~\ref{sec:gene_qe}, where these annotations can be appended to the query when the relevant gene mention is unambiguously resolved. 
This expansion setting is available for 563 queries.

\paragraph{Evidence labels and full-text subset.}
Curator citations define article-level relevance, but they do not guarantee that 
the cited abstract contains the specific evidence needed to support the full 
curator claim. 
For example, in the \emph{pkaC} instance shown in Fig.~\ref{fig:dataset}B, the cited abstract describes the study at a broad level, but not the specific expression pattern stated in the curator claim.
This gap creates an important distinction for retrieval evaluation: 
some queries can be answered from abstracts alone, whereas others 
may require evidence from the article body even when the cited PMID is correct. 
To characterize this distinction, we assigned an evidence-level label to each 
claim--article pair to indicate whether the cited abstract: (i) supports the detailed claim; 
(ii) supports only the core statement; or (iii) is insufficient, suggesting that 
the cited article may require full-text evidence. The third label does not mark 
the citation as irrelevant; rather, it identifies cases where the supporting 
evidence may lie outside the abstract. These labels define the strata for evaluating whether full-text retrieval helps when abstracts provide different levels of evidence. We therefore construct a full-text corpus for articles with accessible PDFs, covering 1,124 cited PubMed articles. Queries with at least one
cited PMID in this corpus form the has-PDF subset, containing 1,480 of the 1,656
queries. Because labels are assigned to claim--article pairs rather than queries,
label counts can exceed the number of queries. In the full dataset,
\emph{abstract supports detail} accounts for 816 pairs (707 in the has-PDF
subset), \emph{abstract supports core} for 654 (585 in the has-PDF subset), and \emph{abstract
insufficient} for 387 (350 in the has-PDF subset).

\FloatBarrier
\section{Retrieval pipeline and variants}
\label{sec:pipeline}

This section defines the retrieval pipeline and the two corpus/query variants evaluated in the experiments.
To make the pipeline easier to follow, we briefly define the main retrieval terms used below. BM25 ranks articles mainly by overlap with the exact words in the query, whereas dense retrieval ranks articles by semantic similarity between learned query and article vectors. Reciprocal Rank Fusion (RRF) combines ranked lists by favoring articles that appear near the top of either list. Reranking reorders an already retrieved candidate list with a more computationally intensive model. Recall@K asks whether a known cited article appears within the top K retrieved candidates, with K denoting the cutoff rank. MRR@K rewards placing the cited article near the top of the final ranking.

\subsection{Two-stage retrieval pipeline}
\label{sec:two_stage_pipeline}
\begin{figure}[!htbp]
\centering
\includegraphicsifexists[width=0.86\linewidth,height=0.28\textheight,keepaspectratio]{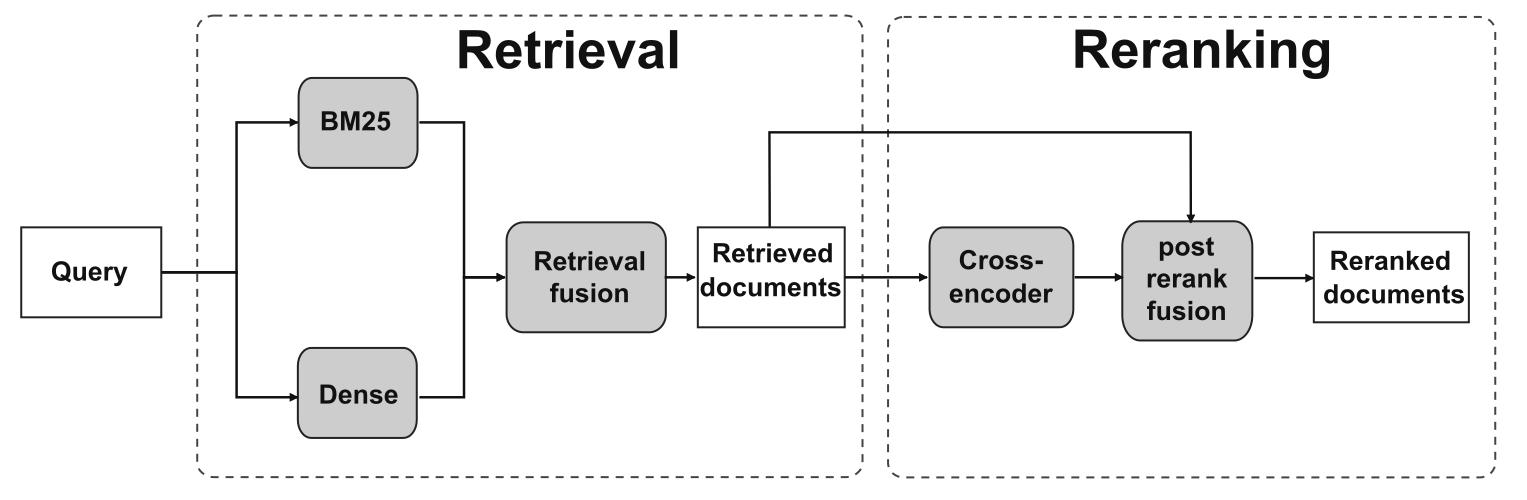}
\caption{Two-stage retrieval and reranking pipeline. 
Stage 1 builds a recall-oriented candidate pool by fusing BM25 and dense retrieval with RRF. 
Stage 2 reranks the fused candidates with a cross-encoder and combines the reranker order with the original retrieval-fusion order through a second RRF step to produce the final ranking.}
\label{fig:pipeline}
\end{figure}

Our pipeline follows a two-stage retrieval-and-reranking design (Fig.~\ref{fig:pipeline}).
Given a query, Stage 1 retrieves candidate articles using two complementary retrievers over the abstract corpus: keyword-based BM25 retrieval and dense semantic retrieval. The two ranked lists are merged with RRF to form a single candidate pool. Stage 2 then reranks this pool with a cross-encoder. The final output combines the cross-encoder ranking with the original retrieval-fusion ranking, retaining a small amount of first-stage signal. The first stage is optimized for high Recall@K at large candidate depths, whereas the reranking stage is evaluated by MRR@K at small K.

\subsection{Gene-aware query expansion}
\label{sec:gene_qe}
We use structured gene annotations from dictyBase to test whether gene-aware query expansion improves retrieval and downstream reranking. For each curator claim, we detect gene mentions using dictyBase gene IDs, canonical names, and synonyms, while filtering short, generic, or highly ambiguous aliases. We compare two expanded query variants with the original claim: (i) + synonyms, which appends the canonical gene name and selected synonyms, and (ii) + synonyms \& products, which additionally appends a short product description. Gene IDs are used only for matching and are not appended to the query. The same expanded query text is used for both retrieval and reranking.

\subsection{Chunked full-text corpus variant}
\label{sec:fulltext_variant}
To test whether evidence beyond abstracts improves retrieval, we compare the abstract-only corpus with a chunked full-text variant, where each article is split into smaller text segments, or chunks, that can be retrieved separately. In the abstract-only setting, each document is one PubMed abstract; in the full-text setting, each article is represented by an abstract chunk and body-text chunks. The same retrieval, fusion, reranking, and post-rerank fusion steps are then applied without changing the pipeline parameters.

Because an article is split into several chunks, the same article can appear multiple times in a chunk-level ranking, while our relevance judgments are defined per article. We therefore collapse the chunk ranking back to articles: each article takes the rank of its highest-ranked chunk. The resulting article-level ranking lets us compare the abstract-only and chunked corpora on the same article-level relevance judgments.

\FloatBarrier
\section{Experimental setup}
\label{sec:experimental_setup}

We evaluate all systems under fixed retrieval depths, fusion settings, and significance-testing procedures. This section defines the metrics and compared systems; implementation details are provided in the Appendix.

\subsection{Metrics, candidate depths, and significance testing}
\label{sec:metrics_significance}
We report Recall@K for the first-stage retrieval and MRR@K for the final reranked output. Because our judgments are derived from inline citations rather than exhaustive relevance assessments, MRR@K is used as the primary final-ranking metric and should be interpreted as measuring retrieval of known cited evidence, not retrieval of all relevant articles.

Pipeline parameters were fixed before the dictyBase experiments and were not tuned on this dataset. We retrieve 5,000 candidates per retriever, fuse the BM25 and dense lists, and pass the top 1,000 fused candidates to the reranker as a fixed, recall-oriented candidate depth. Final rankings are evaluated at small K, primarily MRR@10.

For significance testing, we use paired bootstrap resampling over queries with B = 10,000 samples. We report 95\% percentile confidence intervals for per-query MRR@10 differences against the relevant baseline.

\subsection{Systems compared}
\label{sec:systems_compared}
We compare BM25, BM25+Dense retrieval fusion without reranking, four cross-encoder rerankers, and one external listwise reference system. All cross-encoder rerankers are applied to the same BM25+Dense RRF candidate pool (Fig.~\ref{fig:pipeline}). The rerankers are MiniLM-L12 (cross-encoder/ms-marco-MiniLM-L12-v2)~\cite{wang_minilm_2020,noauthor_cross-encoder-ms-marco-minilm-l12-v2_2026}, BGE-m3 (BAAI/bge-reranker-v2-m3)~\cite{chen_m3-embedding_2024,noauthor_baaibge-reranker-v2-m3_2026}, MedCPT (ncbi/MedCPT-Cross-Encoder)~\cite{noauthor_ncbimedcpt-cross-encoder_2026}, and BGE-Gemma (BAAI/bge-reranker-v2-gemma)~\cite{noauthor_baaibge-reranker-v2-gemma_2026}, representing small general-purpose, strong multilingual, PubMed-trained biomedical, and large LLM-based CE reranking settings.

We also include Ragnarok BM25 + RankZephyr as an external reference~\cite{pradeep_ragnarok_2025,pradeep_rankzephyr_2023}. Ragnarok is an independently maintained retrieval-and-listwise-reranking pipeline using RankZephyr, a 7B listwise LLM reranker. We do not retune Ragnarok on the dictyBase dataset, so this comparison is intended as a contextual reference rather than a parameter-matched baseline.

\section{Results}
\label{sec:results}

We organize the results around three questions: whether cross-encoder reranking
adds value beyond first-stage retrieval baselines, whether curated gene
annotations help compact biological queries, and whether full-text chunks improve
retrieval when abstracts lack sufficient evidence. The aggregate results are reported in Table~\ref{tab:ranker_comparison} and Figs.~\ref{fig:qe}--\ref{fig:evidence_haspdf};
we use short query-level examples within each subsection to illustrate the mechanisms behind these results.

\subsection{Reranking gains are model-dependent}
\label{sec:results_reranking}
On the 1,656-query dataset, BM25 is a strong baseline for top-ranked evidence
retrieval (MRR@10 = 0.593; Table~\ref{tab:ranker_comparison}). BM25+Dense fusion
is used as a recall-oriented candidate generator, but
scores below BM25 at MRR@10, indicating that higher candidate recall does not automatically translate into better top-ranked results. Cross-encoder reranking is also not uniformly beneficial:
the lightweight MiniLM-L12 reranker hurts performance, and BGE-m3 gives
only a small, non-significant gain. Clear improvements appear for the biomedical
MedCPT reranker and the high-capacity BGE-Gemma model, which improve MRR@10 by
+5.7 and +11.4 percentage points, respectively. The external Ragnarok
BM25+RankZephyr reference remains below BM25, further suggesting that gains
depend on reranker fit rather than on adding a reranking stage alone.

\begin{table}[!htbp]
\centering
\caption{Ranker comparison on the full dataset (1,656 queries). Model size denotes the approximate number of parameters for neural rerankers; ``--'' indicates systems without a neural reranker. MRR@10 is the mean reciprocal rank at cutoff 10, where higher values indicate better top-rank performance. $\Delta$ vs.\ BM25 reports the difference in MRR@10 from the BM25 baseline in percentage points; positive values indicate improvement over BM25, and brackets show 95\% paired-bootstrap confidence intervals.}
\label{tab:ranker_comparison}
\scriptsize
\setlength{\tabcolsep}{3pt}
\begin{tabular}{@{}llccc@{}}
\toprule
System & Description & Model size & MRR@10 & $\Delta$ vs.\ BM25 \\
\midrule
BM25       & Lexical baseline              & --          & 0.593 & --                       \\
BM25+Dense & Lexical+dense RRF             & --          & 0.568 & $-2.6$ [$-4.1$,$-1.0$]   \\
MiniLM-L12 & Lightweight MS MARCO CE       & 33M         & 0.541 & $-5.2$ [$-7.0$,$-3.4$]   \\
BGE-m3     & Strong multilingual CE        & 0.6B        & 0.605 & $+1.2$ [$-0.5$,$+2.8$]   \\
MedCPT     & PubMed-trained biomedical CE  & $\sim$110M  & 0.650 & $+5.7$ [$+4.2$,$+7.2$]   \\
BGE-Gemma  & LLM-based CE reranker         & 3B          & 0.707 & $+11.4$ [$+9.8$,$+13.0$] \\
Ragnarok   & External BM25+RankZephyr      & 7B          & 0.566 & $-2.7$ [$-4.8$,$-0.8$]   \\
\bottomrule
\end{tabular}
\end{table}
The per-query analysis shows the same pattern (Fig.~\ref{fig:s1}): MiniLM-L12 hurts more
queries than it helps, BGE-m3 is nearly balanced, and MedCPT and BGE-Gemma shift
more queries toward positive deltas. A concrete failure case is the query
``\emph{DymA} localizes to the phagosome and the cleavage furrow.'' BM25 ranked the
cited article second, whereas MiniLM-L12 moved it down to 60th. The article
ranked first by MiniLM-L12 matched broad location terms, such as cleavage furrow
and phagocytic cup, but lacked the \emph{DymA}-specific signal. Thus, reranking condensed
biological statements is useful only when the reranker preserves the task-relevant
lexical signal.

\subsection{Gene-aware query expansion improves retrieval and benefits BGE-m3 and MedCPT}
\label{sec:results_qe}
We next evaluate gene-aware query expansion on the 563-query subset for which dictyBase synonym and product fields are available. We compare the original query (the curator claim) with two expanded variants: adding gene synonyms, and adding both synonyms and product descriptions.

\begin{figure}[!htbp]
\centering
\includegraphicsifexists[width=0.98\linewidth,height=0.3\textheight,keepaspectratio]{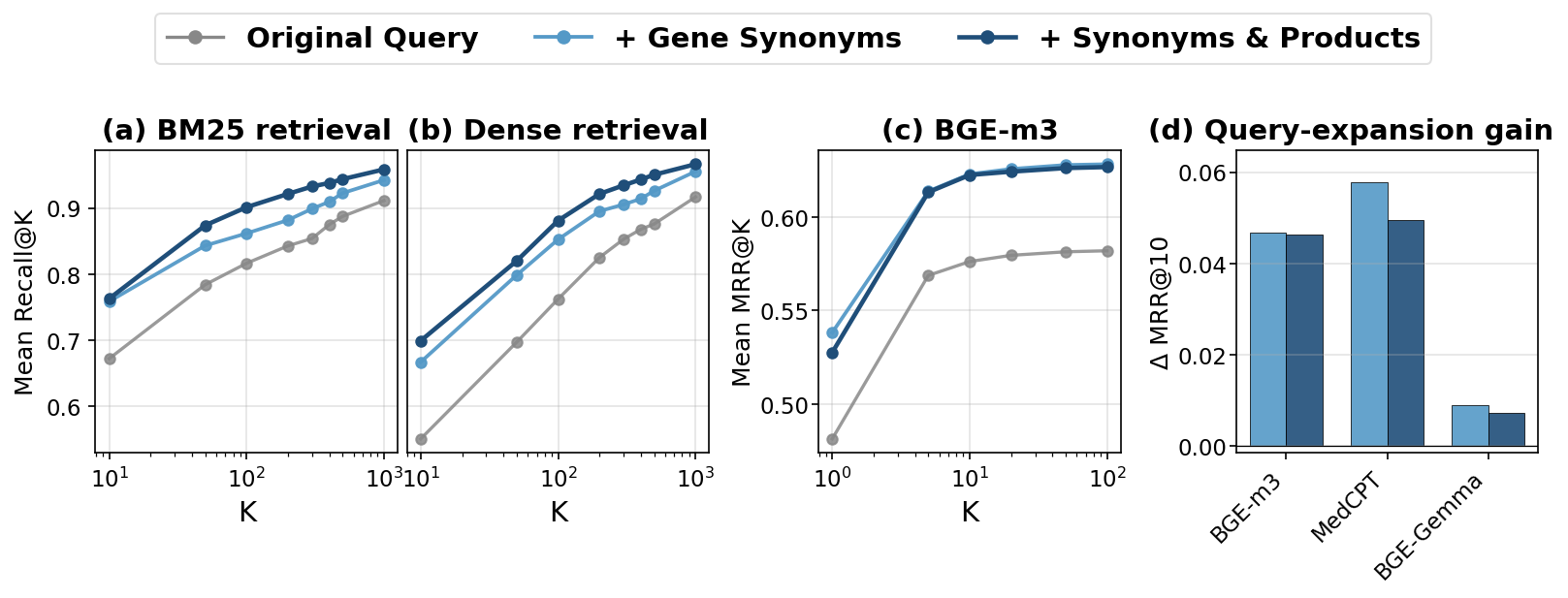}
\caption{Effect of gene-aware query expansion. Three query variants are compared: the original query (the curator claim), the claim plus gene synonyms, and the claim plus both gene synonyms and product descriptions. (a) BM25 Recall@K, (b) dense Recall@K, and (c) mean MRR@K after BGE-m3 reranking, each plotted against the cutoff rank K on the x-axis. (d) Change in MRR@10 (y-axis, in MRR@10 units) relative to the original-query baseline for BGE-m3, MedCPT, and BGE-Gemma; the two bars per reranker are the two expansion variants, and positive values indicate that expansion improved the final ranking. Paired-bootstrap confidence intervals for the MRR@10 deltas are reported in Table~\ref{tab:s1_qe_reranker}.}
\label{fig:qe}
\end{figure}

At the retrieval stage, expansion substantially improves recall for both BM25 and dense retrieval (Fig.~\ref{fig:qe}a--b). For the reranking analysis, we fix the candidate pool to the one retrieved with the deepest expansion (+ synonyms \& products), and vary only the query text passed to the reranker, isolating the query-formulation effect from the recall improvement above. Expanded reranker queries improve MRR@10 for BGE-m3 and MedCPT under both expansion variants, with all four confidence intervals above zero (Fig.~\ref{fig:qe}c--d, Table~\ref{tab:s1_qe_reranker}). By contrast, BGE-Gemma already performs strongly with the original query, and expansion adds less than one point (CIs cross zero).

Overall, gene-aware expansion improves first-stage recall and provides clear downstream benefit for BGE-m3 and MedCPT. A clear example is the query ``\emph{DetA} is involved in cell-type specification during development.'' The cited article uses the synonym \emph{DET1} rather than \emph{DetA}. In the controlled reranking setting, adding this curated synonym moved the cited article from 39th to first under BGE-m3. The synonym supplies the surface form the article actually uses, closing the vocabulary gap between the compact curator wording and the article text. The effect is much smaller for BGE-Gemma, indicating diminishing returns when the reranker already recovers much of the relevant signal from the original query.

\FloatBarrier

\subsection{Full-text chunks benefit claims whose evidence is missing from the abstracts}
\label{sec:results_fulltext}
The previous experiments used the abstract corpus. This setting is limiting for claims where the cited abstracts do not contain sufficient evidence. We therefore evaluate the same retrieval-and-reranking pipeline on the chunked full-text corpus variant (Section~\ref{sec:fulltext_variant}).

\begin{figure}[!htbp]
\centering
\includegraphicsifexists[width=0.92\textwidth,keepaspectratio]{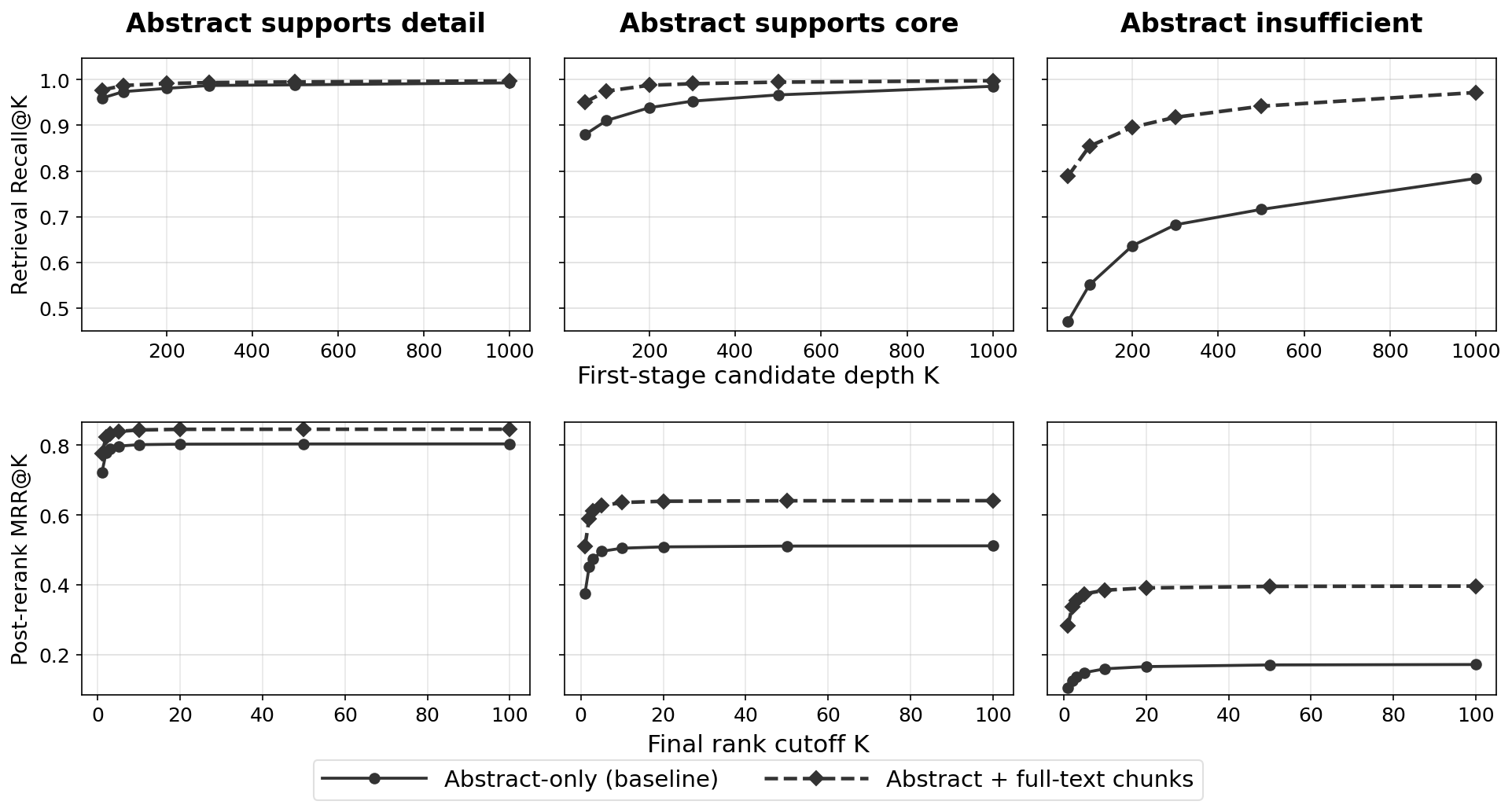}
\caption{Retrieval and reranking by abstract-evidence level on the has-PDF subset. The upper row shows first-stage Recall@K from BM25+Dense RRF retrieval. The lower row shows post-rerank MRR@K using BGE-m3. Columns correspond to \emph{abstract supports detail} (n=707), \emph{abstract supports core} (n=585), and \emph{abstract insufficient} (n=350). Solid lines show the abstract-only baseline; dashed lines show the corpus augmented with full-text chunks.}
\label{fig:evidence_haspdf}
\end{figure}

We stratify queries by the evidence-level labels introduced in the Dataset section: \emph{abstract supports detail}, \emph{abstract supports core}, and \emph{abstract insufficient}. Fig.~\ref{fig:evidence_haspdf} reports the main comparison on the has-PDF subset, and Table~\ref{tab:s2_evidence_haspdf_full} extends it to the full dataset, where full-text coverage is partial.

The main effect of full-text chunks is at the retrieval stage. For abstract-insufficient claims, first-stage Recall@1000 rises from 0.78 to 0.97 (Fig.~\ref{fig:evidence_haspdf}, upper panel), nearly closing the gap to the two abstract-supported groups. The other two groups were already close to the ceiling under the abstract-only corpus and therefore show only small recall gains. Thus, when the cited abstract lacks the required evidence, adding full-text chunks substantially improves whether the cited PMID appears in the candidate pool.

The ranking stage also improves, although the evidence-level gap remains (Fig.~\ref{fig:evidence_haspdf}, lower panel). With BGE-m3, post-rerank MRR@10 on abstract-insufficient claims more than doubles from 0.16 to 0.39, yet stays well below the two abstract-supported groups under the same chunked setting.

At the per-query level, chunked full-text retrieval improved many more abstract-insufficient queries than it hurt (46.6\% improved, 8.0\% decreased). A typical case is the query ``\emph{gefE} null mutants are less sensitive to DIF than wild type.'' In the abstract-only corpus, the cited article ranked 54th because the abstract gives broad developmental context but omits the specific DIF-sensitivity evidence. With full-text chunks, the cited article ranked first because the relevant evidence appears in the article body.

Overall, full-text chunks provide the largest benefit on claims whose cited abstracts lack sufficient evidence. They nearly close the first-stage recall gap and more than double post-rerank MRR@10 for the abstract-insufficient group. The remaining gap to the abstract-supported groups indicates that full-text retrieval is not only a candidate-generation problem: after relevant papers enter the pool, ranking them among chunk-derived candidates remains more difficult.

\FloatBarrier
\section{Discussion}
\label{sec:discussion}
We built and released a benchmark for \emph{Dictyostelium} literature retrieval that reflects a realistic form of a niche biological search: finding supporting citations for compact, gene-centered biological statements. The benchmark's value is a combination of a new dataset and a test case for retrieval behavior that can differ from many open-domain QA settings. In this model-organism setting, gene symbols, pathway terms, phenotypes, and other organism-specific terms provide highly informative lexical cues. These cues help explain why BM25 remains strong and why reranking helps only when the model preserves these signals while adding useful semantic discrimination.

This finding has a practical implication for standard two-stage retrieval systems: cross-encoder reranking should not be treated as a default upgrade to the retrieval pipeline. In niche biological search, model choice, latency, and domain fit should be evaluated together. A reranking stage that adds substantial computation is useful only if it improves over a strong lexical baseline on the target retrieval task. 

Gene-aware query expansion provides a complementary, low-overhead way to improve the pipeline.
In our controlled reranking analysis, we fix the candidate pool and vary only the
reranker query, isolating a second route by which expansion helps: beyond
retrieving more candidates at the first stage, it also gives the reranker better
query-side context. These gains suggest that curated gene annotations can enrich
compact claims with missing gene vocabulary without substantially increasing
reranking noise. More broadly, trusted gene annotations plus a smaller domain-suitable
reranker may be a more efficient operating point than relying only on larger
LLM-based rerankers.

The full-text experiments show that abstracts are efficient but incomplete
retrieval units. Body chunks help most when cited abstracts are insufficient, yet
the remaining MRR gap indicates that full text also creates a harder ranking
problem over noisier, more fine-grained candidates. In practice, not every article has open-access, machine-readable full text, so a
system cannot assume complete full-text coverage. This partial coverage could in
principle bias retrieval, because full-text-covered articles contribute many chunks
whereas abstract-only articles contribute only one. However, the full-dataset columns
of Table~\ref{tab:s2_evidence_haspdf_full} suggest that this effect is not dominant:
mixed abstract/full-text retrieval still improves evidence finding under partial
coverage.

Several limitations remain. The queries were automatically extracted, cleaned, and deduplicated from dictyBase pages using rule-based procedures rather than being rewritten or reviewed in detail by domain experts. Some compact statements may therefore depend on surrounding page or paragraph context and be less self-contained than manually written search queries. In addition, inline citations provide useful retrieval targets but not exhaustive or definitive relevance judgments: other relevant articles may exist but remain unlabeled, and some difficult cases may reflect claim--citation mismatch rather than retrieval failure. For example, an inline citation may support broader page context, may require context lost during claim extraction, or may occasionally point to a paper that does not clearly support the extracted statement. The benchmark is therefore best interpreted as a cited-article retrieval task rather than a complete relevance benchmark. Finally, the evidence-level labels are LLM-assisted analysis labels, not definitive expert annotations. Future work should extend the benchmark with expert-reviewed queries, broader model-organism coverage, and evidence-level evaluation of generated answers.

\FloatBarrier
\section*{Appendix}
\FloatBarrier
\setcounter{figure}{0}
\renewcommand{\thefigure}{S\arabic{figure}}
\renewcommand{\theHfigure}{S\arabic{figure}}
\setcounter{table}{0}
\renewcommand{\thetable}{S\arabic{table}}
\renewcommand{\theHtable}{S\arabic{table}}

\subsection*{Implementation details}

BM25 retrieval uses PyTerrier~\cite{macdonald_pyterrier_2020} over a Terrier index with the default BM25 weighting model. Dense retrieval uses abhinand/MedEmbed-small-v0.1~\cite{noauthor_abhinandmedembed-small-v01_nodate} to encode queries and documents, with cosine similarity over an HNSW index. Retrieval fusion uses RRF with $k=60$ and equal weights for BM25 and dense retrieval. Cross-encoder rerankers receive the top 1,000 fused candidates and are run with \texttt{max\_length}=512. Final rankings are produced by post-rerank fusion: RRF combines the reranker top 200 with the retrieval-fusion top 200 using weights 0.8 and 0.2, respectively. These settings keep the final ranking primarily reranker-driven while preserving part of the retrieval-stage signal.
Evidence-level labels are produced once using a fixed Llama-3.3 prompt schema~\cite{grattafiori_llama_2024} and frozen for all downstream analyses. Code, workflow configurations, and the public dataset (released as the dictyBase citation goldset) are available at \url{https://github.com/fulaibaowang/dictycite}; the repository also includes the reusable retrieval/reranking pipeline used in the experiments. Variants of this pipeline have also been used in BioASQ Task 14B submissions, but the conclusions reported here are based on the dictyBase benchmark and ablation analyses. The Ragnarok comparison uses code from \url{https://github.com/castorini/ragnarok}.

\begin{figure}[H]
\centering
\includegraphicsifexists[width=0.88\linewidth,height=0.34\textheight,keepaspectratio]{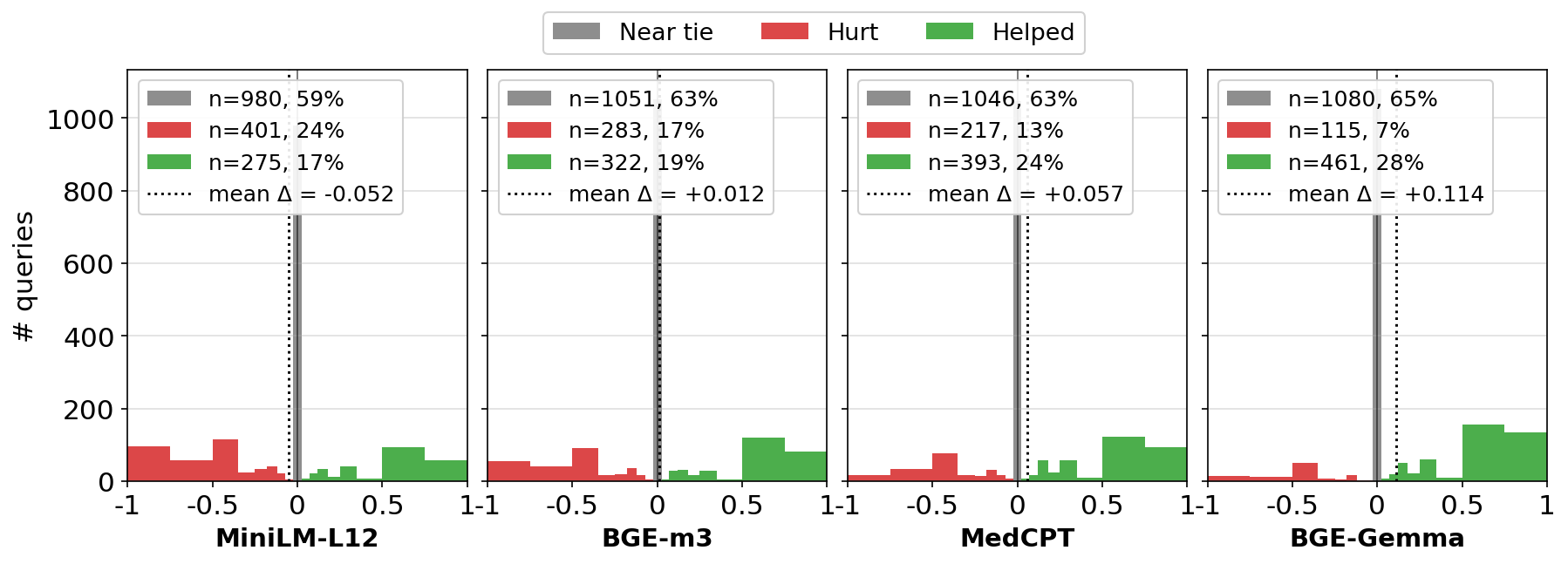}
\caption{Per-query reranking effect relative to BM25. $\Delta$MRR@10 is computed as reranker minus BM25 and grouped as hurt, helped, or near tie using thresholds of $\pm0.025$. Similar aggregate scores mask different query-level behavior: MiniLM-L12 is mostly harmful, BGE-m3 is balanced, and MedCPT and BGE-Gemma shift more queries toward positive deltas.}
\label{fig:s1}
\end{figure}


\FloatBarrier

\begin{table}[H]
\centering
\caption{Effect of gene-aware query expansion on the query-expansion subset. Each cell reports MRR@10. For expanded-query variants, parentheses report the paired-bootstrap difference from the original-query baseline for the same reranker, in percentage points, with 95\% confidence intervals.}
\label{tab:s1_qe_reranker}
\scriptsize
\setlength{\tabcolsep}{4pt}
\begin{tabular}{@{}l c c c@{}}
\toprule
Reranker & Original & + Synonyms & + Synonyms and products \\
\midrule
BGE-m3 & 0.576 & 0.623 ($+4.7$, [$+2.6$, $+6.8$]) & 0.622 ($+4.6$, [$+2.2$, $+7.1$]) \\
MedCPT & 0.624 & 0.682 ($+5.8$, [$+3.7$, $+7.9$]) & 0.673 ($+5.0$, [$+2.7$, $+7.2$]) \\
BGE-Gemma & 0.694 & 0.703 ($+0.9$, [$-0.5$, $+2.3$])$^{\dag}$ & 0.701 ($+0.7$, [$-0.6$, $+2.1$])$^{\dag}$ \\
\bottomrule
\end{tabular}

\vspace{1mm}
{\scriptsize $^{\dag}$ Confidence interval crosses zero. Deltas are in percentage points.}
\end{table}

\begin{table}[H]
\centering
\caption{First-stage Recall@1000 and post-rerank MRR@10 (BGE-m3) by
abstract-evidence level, shown for the has-PDF subset
(Fig.~\ref{fig:evidence_haspdf}) and for the full 1,656-query dataset with
partial full-text coverage. Full-dataset gains are diluted because some cited
PMIDs lack chunked-corpus coverage, but the qualitative pattern is unchanged.}
\label{tab:s2_evidence_haspdf_full}
\scriptsize
\setlength{\tabcolsep}{4pt}
\resizebox{\textwidth}{!}{%
\begin{tabular}{@{}l rcccc rcccc@{}}
\toprule
 & \multicolumn{5}{c}{has-PDF subset} & \multicolumn{5}{c}{full dataset} \\
\cmidrule(lr){2-6}\cmidrule(l){7-11}
 & & \multicolumn{2}{c}{R@1000} & \multicolumn{2}{c}{MRR@10}
 & & \multicolumn{2}{c}{R@1000} & \multicolumn{2}{c}{MRR@10} \\
\cmidrule(lr){3-4}\cmidrule(lr){5-6}\cmidrule(lr){8-9}\cmidrule(l){10-11}
Evidence level & $n$ & Abs & +Chk & Abs & +Chk & $n$ & Abs & +Chk & Abs & +Chk \\
\midrule
\emph{Abstract supports detail} & 707 & 0.993 & 0.997 & 0.802 & 0.843 & 816 & 0.993 & 0.997 & 0.803 & 0.829 \\
\emph{Abstract supports core}   & 585 & 0.986 & 0.998 & 0.505 & 0.636 & 654 & 0.982 & 0.992 & 0.503 & 0.608 \\
\emph{Abstract insufficient}    & 350 & 0.784 & 0.972 & 0.161 & 0.385 & 387 & 0.788 & 0.949 & 0.164 & 0.357 \\
\bottomrule
\end{tabular}%
}
\end{table}

\begin{credits}
\subsubsection{\ackname} This publication is co-funded by the European Union's Horizon Europe research and innovation program under the Marie Sk\l{}odowska-Curie COFUND Postdoctoral Programme grant agreement No.~101081355--SMASH, and by the Republic of Slovenia and the European Union from the European Regional Development Fund. We thank the FRIDA cluster at the University of Ljubljana and HPC Vega at IZUM for providing the computational resources used in this work.

\subsubsection{Disclaimer.} Co-funded by the European Union. Views and opinions expressed are however those of the author(s) only and do not necessarily reflect those of the European Union or European Research Executive Agency. Neither the European Union nor the granting authority can be held responsible for them.

\subsubsection{\discintname} The authors have no competing interests to declare that are relevant to the content of this article.
\end{credits}

\bibliographystyle{splncs04}
\bibliography{ds_paper_revised}

\end{document}